\documentclass[twocolumn,showpacs,preprintnumbers,amsmath,amssymb,prl,aps]{revtex4}
\usepackage{graphicx}
\usepackage{natbib}

\usepackage{dcolumn}
\usepackage{amsmath}

\makeatletter
\def\btt#1{\texttt{\@backslashchar#1}}
\DeclareRobustCommand\bblash{\btt{\@backslashchar}}
\makeatother

\usepackage{color}

\begin{document}

\preprint{HEP/123-qed}


\title[Short Title]{Large anisotropy of spin-orbit interaction in a single InAs self-assembled quantum dot}

\author{S. Takahashi$^1$}\email{stakahashi@meso.t.u-tokyo.ac.jp}
\author{R. S. Deacon$^{1, 2}$}
\author{K. Yoshida$^3$}
\author{A. Oiwa$^{1, 2, 3}$}
\author{K. Shibata$^4$}
\author{\\K. Hirakawa$^{2, 4, 5}$}
\author{Y. Tokura$^{2, 6}$}\email{tokura@will.brl.ntt.co.jp} 
\author{S. Tarucha$^{1, 3, 5}$}\email{tarucha@ap.t.u-tokyo.ac.jp}

\affiliation{
$^1$Department of Applied Physics and QPEC, University of Tokyo, 7-3-1 Hongo, Bunkyo-ku, Tokyo 113-8656, Japan\\
$^2$JST CREST, 4-1-8 Hon-cho, Kawaguchi-shi, Saitama 332-0012, Japan\\
$^3$ICORP JST, 3-1 Wakamiya, Morinosato, Atsugi-shi, Kanagawa 243-0198, Japan\\
$^4$IIS, University of Tokyo, 4-6-1 Komaba, Meguro-ku, Tokyo 153-8505, Japan\\
$^5$INQIE, University of Tokyo, 4-6-1 Komaba, Meguro-ku, Tokyo 153-8505, Japan\\
$^6$NTT basic research laboratry, 3-1 Wakamiya, Morinosato, Atsugi-shi, Kanagawa 243-0198, Japan
}


\date{\today}


\begin{abstract}
Anisotropy of spin-orbit interaction (SOI) is studied for a single uncapped InAs self-assembled quantum dot (SAQD) holding just a few electrons.
The SOI energy is evaluated from anti-crossing or SOI induced hybridization between the ground and excited states with opposite spins.
The magnetic angular dependence of the SOI energy falls on an absolute cosine function for azimuthal rotation, and a cosine-like function for tilting rotation.
The SOI energy is even quenched at a specific rotation.
These angular dependence compare well to calculation of Rashba SOI in a two-dimensional harmonic potential.
\end{abstract}



\maketitle


Spin-orbit interaction (SOI) arises from the magnetic moment of electron-spin coupling to its orbital degree of freedom, and can be significantly modified by the crystal orientation, internal electric field, and quantum confinement effect in semiconductor nanostructures\cite{book:Winkler}.
This feature paves the way to both electrically and magnetically control the spin effect, leading to the novel concept of spin-dependent electronics or spintronics and spin-based quantum information processing\cite{book:Awschalom,book:Maekawa}.
For example, spin transistors\cite{paper:Datta}, and spin qubits\cite{paper:Rashba, paper:Golovach} are based on precession of electron spins about an electrically-induced local magnetic field in semiconductor heterostructures\cite{paper:Nitta}, and semiconductor quantum dots (QDs)\cite{paper:Katja}, respectively.
These SOI effects have been intensively studied for GaAs-based nanostructures.
In general, SOI is strong in narrow gap semiconductors, but it has only recently been studied for InAs and InSb based nanowire QDs\cite{paper:Fasth, paper:Nilsson}.


Although SOI should have anisotropy due to crystal orientation (Dresselhaus term) and asymmetric confinement (Rashba term), there are few reports on these effects\cite{paper:Ganichev,paper:Kohda} and to date no experiment for QDs.
In this work, we used a large uncapped InAs self-assembled QD (SAQD), which has highly anisotropic confinement, to investigate the angular dependence of SOI energy on magnetic field rotation in two directions: tilting rotation out of plane to the sample surface and azimuthal rotation in plane to the surface.
We first measured excitation spectra evolving with magnetic field for a single QD to identify the $S_z = \pm 1/2$ states in different orbitals which can be hybridized by SOI when degenerated with magnetic field.
The SOI energy was evaluated from the size of the anti-crossing between these SOI hybridized states.
We also measured the angular dependence of $g$-factor and compared with that of SOI energy.
We found the $g$-factor reflects the confinement anisotropy, while the SOI energy depends on the direction of SOI induced magnetic field.


InAs SAQDs were grown with Stranski-Krastanov mode by molecular beam epitaxy on a (001) semi-insulating GaAs substrate.
The SAQDs are uncapped on an i-GaAs layer, and an n-doped GaAs layer is embedded with an i-AlGaAs layer above as a current blocking layer.
Gate voltage, $V_g$, was applied to the n-doped GaAs layer to change the number $N$ of electrons in the QD.
A pair of Ti (5 nm)/Au (30 nm) electrodes separated by a 60 nm gap was directly placed on the SAQD, using electron beam lithography techniques (inset in Fig.1(b)).
The paired electrodes are the source and drain contacts to the QD.
The SAQDs have some variations in the lateral size, and we selected a large SAQD with diameter $\sim$100 nm to fabricate the device.
The confining potential in the QD we evaluated has an elliptic shape elongated in $y$-direction probably because of the contact to the electrode metal.
Electron transport was measured in a $^3$He-$^4$He dilution refrigerator equipped with a standard superconducting magnet and an {\it in-situ} single-directional sample rotation.
The refrigerator base temperature is 40 mK.
We rotated the sample to measure the magnetic angular dependence of SOI.
Here, we assume this sample rotation is equivalent to the magnetic field rotation with the coordinates fixed on the sample.


\begin{figure*}[bthp]
\includegraphics[width=0.9\linewidth]{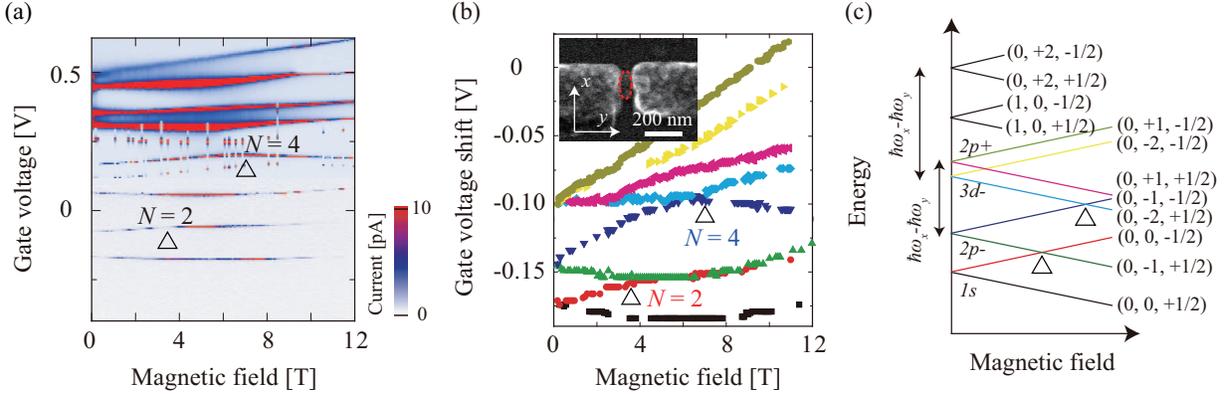}
\caption{\label{fig:ground}
(a) Magnetic evolution of Coulomb peaks measured for $V_{sd} = 50$ $\mu$V.
A series of dots below and above the fourth peak ($N=4$) are due to instability induced by unintentional charging of nearby QD(s).
(b) Peak positions after subtracting the contribution of charging energy from the Coulomb peaks in (a).
Inset: scanning electron microscope image of the device.
An anisotropic QD marked by the dotted line with a height $\approx$ 20 nm bridges the two contact leads.
(c) Schematic of single-particle state energy vs. magnetic field for a 2D QD with the elliptic confinement characterized by $\omega_x$ and $\omega_y$ ($\geq \omega_x$).
Each state is indexed with $(n, l, S_z)$.
The ground state transition subject to SOI occurs for the peaks or states labeled $\triangle$ in each figure.
}
\end{figure*}


Figure 1(a) shows the evolution of Coulomb peaks with perpendicular magnetic field, $B_{\perp}$.
No peaks are observed for $V_g \leq -0.2$ V, indicating the SAQD is empty.
The magnetic evolution of Coulomb peaks reflects the electro-magnetic confinement, Zeeman splitting, and SOI, which lead to a variety of ground state transitions.
We initialy studied ground state transitions signified by upward kinks labeled $\triangle$ for the second ($N=2$) and fourth ($N=4$) Coulomb peaks\cite{paper:Tarucha}.
We subtracted the contributions of charging energy from the Coulomb peaks in Fig.1(a), and plotted the peak positions in Fig.1(b).
Assuming a 2D harmonic potential for the lateral confinement of the QD, states are assigned with radial quantum number $n$, orbital angular momentum $l$ and $z$-component of spin angular momentum $S_z$ using the identifier $(n, l, S_z)$\cite{paper:Igarashi,paper:Tarucha,paper:Jung}.
The lowest two peaks are assigned to the $1s$ Zeeman substates with $(0, 0, \pm1/2)$, and the subsequent two peaks to the $2p_-$ Zeeman substates with $(0, -1, \pm1/2)$.
The next is the $2p_+$ state with $(0, +1, \pm1/2)$, which should be degenerate with the $2p_-$ state.
However, this is not the case in Fig.1(b), due to the anisotropy in the lateral confinement.
We characterized the anisotropy using two lateral confinement components, $\hbar \omega_x = 4$ meV and $\hbar \omega_y = 1.5$ meV.
Also, because of the anisotropy, a $3d_-$ state with $(0, -2, \pm1/2)$ is close to the $2p_+$ state.


\begin{figure*}[bthp]
\includegraphics[width=0.95\linewidth]{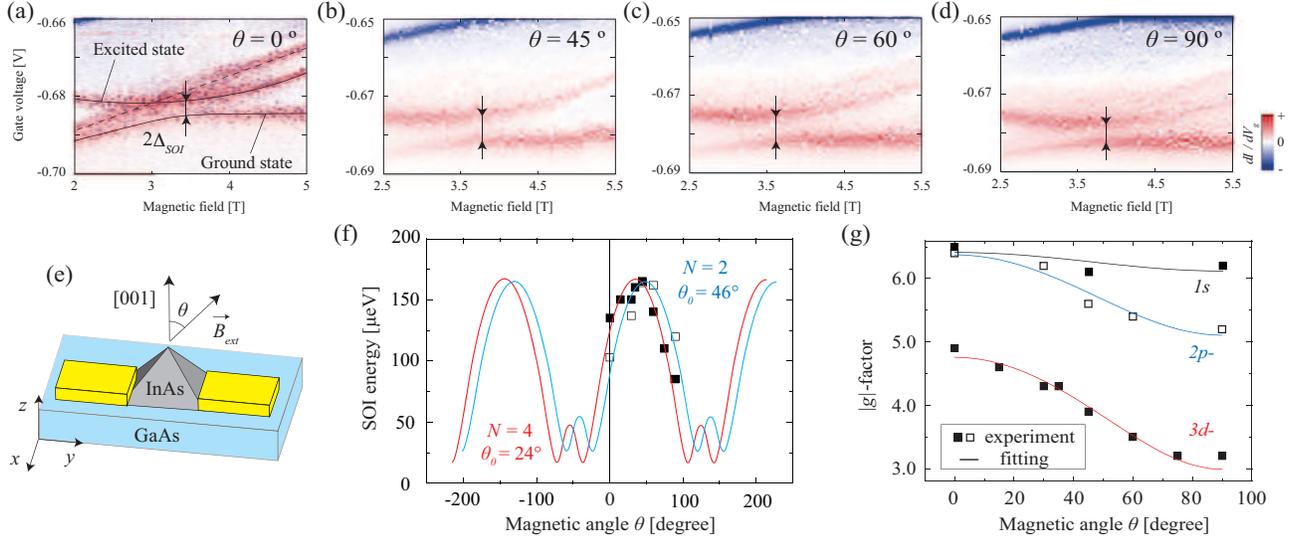}
\caption{\label{fig:theta}
$\underline{Tilting Rotation}$ $dI/dV_g$ vs. $B_{ext}$ - $V_g$ near the $N=2$ ground state transition measured for $V_{sd} = 1.5$ mV at various tilting angles $\theta$ (See (e)): $\theta = 0^\circ$(a), $45^\circ$(b), $60^\circ$(c), $90^\circ$(d), respectively.
In each figure, the lowest line in red indicates the ground state, and the excited state is identified by the line in red located above.
Note an unknown excited state appears in (a) (explained in text).
(f) The SOI energy vs. $\theta$ for the $N=2$ and 4 state derived from the anti-crossings between the ground and excited state.
(g) The $|g|$-factor vs. $\theta$ for the $1s$, $2p_-$, and $3d_-$ state derived from the Zeeman splittings of the $N=1$, 3, 5 states, respectively\cite{EPAPS-SOI}.
The solid line in (f) and (g) are calculations to fit the data points.
}
\end{figure*}


We focus on the two peak crossings labeled $\triangle$ in Fig.1(b) or $N=2$ and 4 ground state transition labeled $\triangle$ in Fig.1(a) to explore the SOI effect.
The angular momentum is a good quantum number under a high $B_{\perp}$ field despite the lateral electrostatic anisotropy.
The $N = 2$ state at $B_{\perp} = 3.5$ T then results from SOI hybridization between the (0, 0, -1/2) and (0, -1, +1/2) state.
The $N = 4$ state at $B_{\perp} = 7.2$ T results from hybridization between the (0, -1, -1/2) and (0, -2, +1/2) state.
These SOI arises from the Rashba term but not the Dresselhaus term, because the former holds for the condition $\Delta l + \Delta S_z = 0$, while the latter for $\Delta l - \Delta S_z = 0$\cite{paper:Destefani}.
We measured the excitation spectra for the $N=2$ and 4 states at around the $\triangle$ points to study the effect of SOI hybridization.
The results for the $N=2$ for a range of tilt angle $\theta$ to the external magnetic field, $\overrightarrow{B}_{ext}$ are shown in Fig.2(b) to (e).
In all figures we observe anti-crossing between the ground and excited state.
The ground and excited state compare well to the $N=2$ and 3 Coulomb peak, so that the anti-crossing is assigned to the SOI hybridization of the (0, 0, -1/2) and (0, -1, -1/2) state.
The anti-crossing size corresponds to twice the SOI energy $\Delta_{SOI}$, and is evaluated to be 70 to 160 $\mu$eV depending on $\theta$ for $N = 2$, which is of the same order as that previously reported on an InAs nanowire QD\cite{paper:Fasth}.
As $\theta$ increases, the $\Delta_{SOI}$ becomes maximal at $\theta \approx 45^\circ$ and then decreases.
Note only in Fig.2(a), we see an additional excited state (dashed line), which crosses with the excited state of interest.
This excited state may reflect a magnetic orbital effect rather than Zeeman effect, because it is absent for the lager $\theta$.
Such a magnetic effect can be associated with the pyramidal QD shape, but does not seem to influence the anti-crossing considered here.
We performed the same measurement about the $N=4$ anti-crossing and identified the SOI hybridization of the (0, -1, -1/2) and (0, -2, +1/2) state\cite{EPAPS-SOI}.
The change of $\Delta_{SOI}$ with $\theta$ obtained for both $N = 2$ and $N = 4$ is shown in Fig.2(f).
The $\theta$-dependence is well reproduced by a cosine-like function, which is discussed later.
$\Delta_{SOI}$ vs. $\theta$ is somewhat displaced between $N=2$ and 4, suggesting that SOI anisotropy slightly depends on $N$.


We also measured the excitation spectra for $B_{\perp} \leq 1$ T for the first, third, and fifth Coulomb peaks to evaluate the $g$-factor for the $1s$, $2p_-$, and $3d_-$ states, respectively\cite{EPAPS-SOI}, and derived $|g| = 6.5$, $6.4$, and $4.9$ for the respective states.
The smaller $|g|$-factor for the higher-lying orbital state probably reflects the weaker coupling between the conduction and valence band.
Using these values, we calculated the spin-orbit length of about 410 nm for our InAs SAQD, which is much smaller than that for GaAs based QDs\cite{paper:Katja, paper:Michel}.


The $\theta$ dependence of the $|g|$-factor is shown for various orbital states in Fig.2(g).
The $|g|$-factor is maximal at $\theta = 0^\circ$ and minimal at $\theta = 90^\circ$.
This angular dependence can be fitted with a phenomenological cosine-like formula\cite{paper:Mayer Alegre}:
\begin{eqnarray}
g = \sqrt{(g_1 cos\xi)^2 + (g_2 sin\xi)^2},
\end{eqnarray}
where $\xi = \theta$ or $\phi$.
$g_1$ and $g_2$ is the $|g|$-factor $\xi = 0^\circ$, and $90^\circ$, respectively, and assigned as ($g_1, g_2$) = (6.4, 6.3), (6.4, 5.2), and (4.7, 3.1) for $1s$, $2p_-$, and $3d_-$-state, respectively.
This result simply reflects that electrons with larger angular momentum experience stronger magnetic confinement, resulting in the stronger angular dependence of $|g|$-factor\cite{paper:Pryor}.


Rotation of the device azimuthally with angle $\phi$ to $\overrightarrow{B}_{ext}$ (See Fig.3(e)) required the warming up of the dilution fridge and remounting of the device.
Though the thermal cycle modifies the characteristic of the QD, we could reproduce the anti-crossing of the $N = 2$ ground and excited state, but not for the $N=4$ state.
We therefore focused on the $N=2$ state hereafter.
Figures 3(a)-(d) show the $N=2$ excitation spectra measured for various $\phi$ values.
As $\phi$ increases from $30^\circ$, the anti-crossing decreases and even disappears at $\phi \approx 60^\circ$, and then reappears and increases.
Note in Figs.3(a)-(d), an additional higher energy excited state is probably identical to the unidentified excited state in Fig.2(a).
The change of $\Delta_{SOI}$ with $\phi$ is shown in Fig.3(f).
The $\phi$-dependence of $\Delta_{SOI}$ can be fitted well with the absolute value of a cosine function with an offset $\phi_0$.
Note we observed a similar $\phi$-dependence of $\Delta_{SOI}$ for $N=10$ but with a quench at $\phi \approx 30^\circ$.
This again means that SOI anisotropy depends on $N$.
Figure 3(g) shows the $|g|$-factor vs. $\phi$ for the $2p_-$ orbital state measured at the same time.
The $|g|$-factor is constant with $\phi$, and can be fitted by equation (1) with $g_1 = g_2 = 4.1$.


The results obtained so far all suggest that electrons are confined by a 2D potential and experience the Rashba effect associated with an out of plane electric field.
Here we examined the 2D potential anisotropy using the $\phi$-dependence of $B_{ext}$ at the anti-crossing\cite{EPAPS-SOI}.
The anti-crossing $B_{ext}$ is largest (smallest) at $\phi = 0^\circ$ ($90^\circ$).
This indicates the lateral confinement is strongest (weakest) in the $y$($x$)-direction, as defined by the electrode metal (See Fig.1(b) inset).


\begin{figure*}[bthp]
\includegraphics[width=1.0\linewidth]{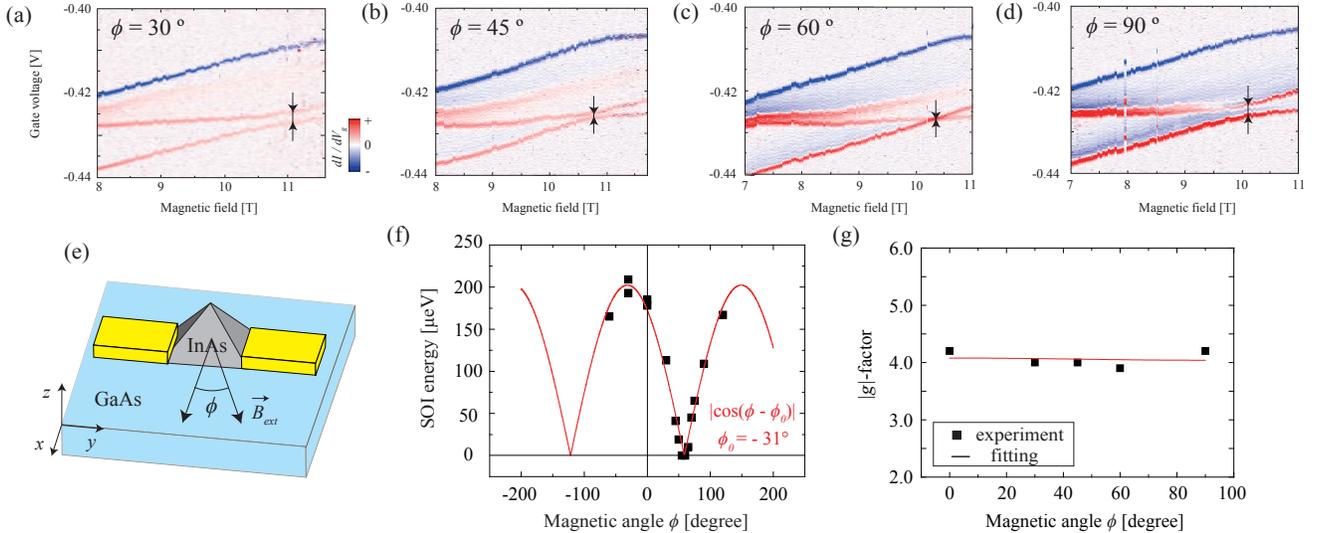}
\caption{\label{fig:phi}
$\underline{Azimuthal Rotation}$ $dI/dV_g$ vs. $B_{ext}$ - $V_g$ near the $N=2$ ground state transition measured for $V_{sd} = 1.5$ mV at various azimuthal angles $\phi$ (See (e)): $\phi = 30^\circ$(a), $45^\circ$(b), $60^\circ$(c), $90^\circ$(d), respectively.
Note an unknown excited state appears in (a)-(d), which is similar to that in Fig.2(a).
(f) The SOI energy vs. $\phi$ for the $N=2$ state derived from the anti-crossings between the ground and excited state.
(g) The $|g|$-factor vs. $\phi$ for the $2p_-$ state derived from the Zeeman splittings of the $N=3$ state.
The solid line in (f) and (g) are calculations to fit the data points.
}
\end{figure*}


Finally, we discuss the Rashba effect on the observed SOI anisotropy.
We assume that the elecric field $\overrightarrow{E}$ caused by the potential gradient in the QD has only a $z$-component: $\overrightarrow{E} = (0, 0, E_z)$.
Theory\cite{paper:Tokura} predicts that the SOI energy for $N=2$ is described as
\begin{eqnarray}
\Delta_{SOI} &=& \mid \langle 1s \downarrow \mid H_{SOI} \mid 2p- \uparrow \rangle \mid \nonumber \\
&=& \mid \langle 1s \downarrow \mid \lambda \overrightarrow{E} \times \overrightarrow{p} \cdot \overrightarrow{\sigma} \mid 2p- \uparrow \rangle \mid \nonumber \\
&=& \lambda E_z \mid (Q_x sin\phi - Q_y cos\phi)cos\theta \nonumber \\
&& \quad -i (E_z Q_x cos\phi + E_z Q_y sin\phi) \mid,
\end{eqnarray}
using the SOI parameter $\lambda$, electron momentum $\overrightarrow{p}$, Pauli matrix vector $\overrightarrow{\sigma}$, and $Q_\nu = \langle 1s \downarrow \mid p_\nu \mid 2p- \uparrow \rangle$ ($\nu = x, y, z$).
Note $Q_z = 0$ in the 2D system.


We first consider the azimuthal rotation ($\phi$), where $\theta = \pi/2$ and $Q_y = 0$ because the $2p_-$ state has a node only in the $x$-direction, so that, 
\begin{eqnarray}
\Delta_{SOI} &=& \lambda E_zQ_x \mid cos\phi \mid.
\end{eqnarray}
This equation is qualitatively consistent with the $\phi$ dependence of $\Delta_{SOI}$, obtained in Fig.3(f).
SOI can only be quenched when $\overrightarrow{B}_{ext}$ and $\overrightarrow{\sigma}$ are aligned.
In the azimuthal rotation, quenching of SOI appears because the Rashba SOI induced $\overrightarrow{E}$ has only a $z$-component and $\overrightarrow{\sigma}$ is always in the $x$-$y$ plane.
The offset $\phi_0 = -31^{\circ}$ could be assigned to anisotropy in the QD potential if there were such anisotropy.
However, we have already found that the QD potential is actually elongated in $x$-direction.
So, we assume spatial displacement of the confining potential between the $1s$ and $2p_-$ state as a possible reason for the finite offset of $\phi_0$, because this gives rise to a finite value of $Q_y$.
In addition, this potential displacement can depend on the specific orbital, accounting for the difference of $\phi_0$ observed between $N=2$ and 10.


For tilting rotation ($\theta$) with $\phi = 0^\circ$, both $Q_x$ and $Q_y$ should depend on $\theta$, because not only the $g$-factor but also the wave function depend on $\theta$ or the perpendicular component of magnetic field.
Therefore, equation (2) becomes more complicated, but the numerical calculation still gives a cosine-like function of $\Delta_{SOI}$ with $\theta$, which fits the experimental data in Fig.2(f).
In this calculation, we arbitrarily set $\phi_0 = -82^\circ$ for both $N=2$ and 4 because $\phi_0$ can be changed after the thermal cycle.
We could not judge whether SOI is quenched with $\theta$ because of the limited $\theta$ range in our experiment.
Note experimental observation of the quenching of SOI in tilting rotation would rarely ocuur.
This is because the quenching only occurs when the magnetic field is exactly aligned with the direction of $\overrightarrow{\sigma}$ which is defined by the potential displacement as already discussed for the $1s$ and $2p_-$ state.
In addition, the theoretical curves in Fig.2(f) include the offsets $\theta_0 = 46^\circ$ and $24^\circ$ for $N = 2$ and $N =4$, respectively.
These angle offsets can also be explained by considering the potential displacement as discussed for the offset $\phi_0$ but in the $z$-direction\cite{paper:Jung}.


Magnetic angular dependence of SOI previously observed for a 2D system\cite{paper:Ganichev} shows a cosine-like feature but no quenching, because both Rashba and Dresslhaus contributions are included.
On the other hand, in our QD by selecting the Rashba effect, SOI can be tuned from zero to a finite value.
This can provide a new scheme of controlling spin manipulation speed depending on magnetic field angle for SOI-induced electron spin resonance.

In summary, SOI anisotropy was investigated for a single uncapped InAs SAQD with magnetic field rotating in and out of plane.
Based on the assignment of the angular and spin quantum numbers, we selected the ground state transitions only associated with the Rashba SOI to evaluate the SOI energy.
The magnetic angular dependence of SOI energy shows good agreement with an absolute cosine function for azimuthal rotation, and with a cosine-like function for tilting rotation, reflecting the direction of Rashba SOI induced field.
On the other hand, the $|g|$-factor only depends on tilting rotation, reflecting the orbital angular momentum of the state.


S. Takahashi is supported by JSPS Research Fellowships for Young Scientists.
This project was supported by Grant-in-Aid for Research S (No. 19104007), A (No. 21244046) and QuEST program (BAR-0824).



\end{document}